\documentclass[twocolumn,aps,prl,10pt,amsmath,amssymb,nofootinbib,showpacs,superscriptaddress,floatfix]{revtex4-1}

\DeclareFontFamily{U}{rcjhbltx}{}
\DeclareFontShape{U}{rcjhbltx}{m}{n}{<->rcjhbltx}{}
\DeclareSymbolFont{hebrewletters}{U}{rcjhbltx}{m}{n}

\usepackage{graphicx}
\usepackage{color}
\usepackage[usenames,dvipsnames]{xcolor}
\usepackage[colorlinks=true,linkcolor=Red,citecolor=Green,linktoc=page]{hyperref}
\usepackage{multirow}
\usepackage{float}
\usepackage{flushend}
\usepackage{balance}
\usepackage[varg]{txfonts}
\usepackage{ulem}
\usepackage{fancyhdr}

\DeclareMathSymbol{\lamed}{\mathord}{hebrewletters}{108}
\begin{document}
\title{Effective de Sitter space, quantum behaviour and large-scale spectral dimension (3+1)}

%\author{C.\,Kelly}

%\affiliation{School of Engineering and Physical Sciences, Heriot-Watt University, Edinburgh EH14 4AS, UK}

\author{C.\,A.\,Trugenberger}

\affiliation{SwissScientific Technologies SA, rue du Rhone 59, CH-1204 Geneva, Switzerland.}

%\author{F.\,Biancalana}

%\affiliation{School of Engineering and Physical Sciences, Heriot-Watt University, Edinburgh EH14 4AS, UK}

%\date{\today}
\begin{abstract}
De Sitter space-time, essentially our own universe, is plagued by problems at the quantum level.  Here we propose that Lorentzian de Sitter space-time is not fundamental but constitutes only an effective description of a more fundamental quantum gravity ground state. This cosmological ground state is a graph, appearing on large scales as a Riemannian manifold of constant negative curvature. We model the behaviour of matter near this equilibrium state as Brownian motion in the effective thermal environment of graph fluctuations, driven by a universal time parameter. We show how negative curvature dynamically induces the asymptotic emergence of relativistic coordinate time and of leading ballistic motion governed by the isometry group of an ``effective Lorentzian manifold" of opposite, positive curvature, i.e. de Sitter space-time: free fall in positive curvature is asymptotically equivalent to the leading behaviour of Brownian motion in negative curvature. The local limit theorem for negative curvature implies that the large-scale spectral dimension of this ``effective de Sitter space-time" is (3+1) independently of its microscopic topological dimension. In the effective description, the sub-leading component of asymptotic Brownian motion becomes Schr\"odinger quantum behavior on a 3D Euclidean manifold.
\end{abstract}
\maketitle

%\section{Introduction}

Our universe is fast approaching de Sitter space-time (for a review see \cite{strominger}). However, a complete and consistent quantum formulation of de Sitter space-time is not devoid of problems \cite{susskind1}. Also, de Sitter space-time is difficult to obtain as a ground state of string theory \cite{dine}. 

But is de Sitter space-time really necessary to describe the fate of our universe? To determine the curvature of space-time one needs to measure the trajectories of test objects. We now show that, asymptotically, the sign of the curvature is ambiguous: ``free fall" in constant positive curvature becomes equivalent to Brownian motion in constant negative curvature. Therefore, in this asymptotic region, it is not possible to determine from the motion of test particles if one is living on de Sitter space-time or on an ``effective de Sitter space-time" which is actually a constant negative-curvature Riemannian manifold with emergent coordinate-time. 

This idea, in addition, may contribute to resolve the issue of the incompatibility of the concept of time in general relativity and quantum mechanics (for a review see \cite{kuchar}). Time, in general relativity and cosmology is associated with the negative signature coordinate of a fundamental Lorentzian manifold. Quantum mechanics, instead is predicated on a Hamiltonian, quantum mechanical time is the corresponding Hamiltonian flow, a concept a priori independent of geometry. 

Recently, we proposed \cite{comb1, comb2, comb3, comb4} a purely combinatorial quantum gravity model based on random graphs governed by the combinatorial Ollivier-Ricci curvature \cite{olli0, olli1, olli2, olli3, olli4}. In this model there is no a priori notion of either time or space, only abstract binary relations, as suggested originally by Wheeler \cite{wheeler} (for a review see \cite{deutsch}). Nevertheless, when graph edges are endowed with a length scale, geometry emerges from pure randomness in a phase transition from strong to weak gravity for which there is strong numerical evidence of continuity \cite{comb2, comb3}. In the geometric phase, ground state graphs are discretizations of negative-curvature Riemannian manifolds \cite{comb3}, so-called Cartan-Hadamard manifolds (see e.g.  \cite{shiga}) \footnote{Strictly speaking, computational limitations have permitted numerical studies only in 2D}. 

Based on this model, here we propose that de Sitter space-time is not fundamental but only an effective asymptotic description of our universe. The cosmological ground state is, rather a graph appearing, on large distances, as a negative-curvature Riemannian manifold. We model the behaviour of matter particles near this equilibrium state by their interaction with the effective ``thermal" environment given by graph fluctuations. Of course, as soon as small deviations from equilibrium are considered, we have to introduce a universal time parameter driving these deviations. We will identify this as the original quantum mechanical time concept. Note that this time has nothing to do with geometry, it is an absolute concept that can be considered as a discrete ordinal variable or represented itself by a graph \cite{pessa}. 

The result of environmental thermal fluctuations on matter particles is Brownian motion (for a review see \cite{eastman}). On flat manifolds this is simply caused by scattering events randomly pushing the particle around. On curved manifolds, however the interplay of scattering with curvature makes Brownian motion look very different \cite{kleinertcur}. This is particularly true for negative-curvature manifolds \cite{ledrappierbal,hsu, hsubrief, arnaudon, hsubook}. 

Here we show how negative curvature forces the leading behaviour of Brownian motion to become asymptotically deterministic and ballistic on the geodesics of the corresponding de Sitter space-time of reversed-sign, positive curvature, playing the role of the cosmological constant. Correspondingly, the probability density of the particle distribution is governed by the wave equation on this ``effective de Sitter space-time", after the universal time is dynamically soldered to a manifold coordinate by the diffusion equation itself. 
Essentially, Brownian scattering events in this emerging coordinate time are asymptotically ``diluted away", so that only the ballistic motion on times scales shorter than the Brownian momentum relaxation time survives. Since time is a manifold coordinate, in this asymptotic region, the admitted ballistic evolutions must be representations of the isometry group ${\rm SO}_+(D,1) $ of the ground state manifold. This is the de Sitter group, which replaces the Poincar\'e group in presence of a cosmological constant \cite{dSrel1, dSrel2, dSrel3}, up to the absence of time-reversal symmetry, which amounts to an emergent arrow of time. The Lorentzian de Sitter picture with the general-relativistic time as a manifold coordinate emerges thus as an asymptotic, effective description of a more fundamental physics. 

A manifold of constant negative curvature is homogeneous and, therefore, any two points can be related by an isometry. This property, however, is lost when there is matter diffusing on the manifold. Indeed, the origin of the Brownian motion becomes a preferred point on the manifold. By an isometry we can always choose the origin of dynamics to be also the origin of our coordinate system. With respect to this common origin, the manifold separates into two regions: in the first region, fluctuations, as seen by Brownian motion, are strong while in the second they are diluted away. As a consequence, the behaviour of these fluctuations with respect to the chosen origin gives an absolute physical meaning to the radial coordinate of the hyperboloid model, see below. In the former region of strong fluctuations there is no general relativity, nor Hamiltonian dynamics, only Brownian diffusion. In the asymptotic region of diluted fluctuations, coordinate time emerges dynamically, motion becomes ballistic and the concept of particles as Lorentz group representations appears. Admittedly, in the present version of the model we have not derived general relativity in this region, although the absolute value of the curvature does appear as a particle mass there.

Brownian motion, generated by the Laplace operator, defines the spectral dimension of a generic metric space, be it a manifold, a fractal or a graph. Essentially, the diffusion process probes the geometric characteristics on a given scale and determines what is the ``effective number of directions" seen by the random walker on that scale. On a Euclidean manifold it coincides, of course, with the topological dimension. On a fractal, however, it must not even be an integer. And negative curvature can have a very large effect too. Due to the leading ballistic character of asymptotic Brownian motion, the diffusion process is far too fast to probe the large-scale geometry. However, 
the sub-leading random component of Brownian motion can still probe the large-scale spectral properties of the negative-curvature manifold \cite{anker}. The resulting spectral dimension is always 3, independent of the topological dimension of the manifold. This result is a general consequence of the local limit theorem for negative-curvature manifolds \cite{ledrappier} and implies that the large-scale spectral dimension of effective de Sitter space-time is always (3+1), independently of its microscopic topological dimension. In the effective Lorentzian description, the asymptotic sub-leading random process becomes Schr\"odinger quantum behaviour on a 3D Euclidean manifold. 
 
Let us consider a Riemannian manifold of constant negative curvature $-H^2$ \cite{shiga}. In polar coordinates \cite{hsu, hsubrief, arnaudon, hsubook} the metric is given by 
\begin{eqnarray}
ds^2 &&= dz^2 + G^2(z) d\Omega^2 \ ,
\nonumber \\
G(z) &&= {{\rm sinh} Hz\over H} \ .
\label{radial}
\end{eqnarray} 
with $d\Omega ^2 $ the standard metric on the (D-1)-dimensional sphere ${\cal S}^{D-1}$, which is conveniently parametrized by $(D-1)$ angles,
\begin{eqnarray}
\omega_1 &&= {\rm cos} \ \theta_1 \ ,
\nonumber \\
\omega_2 &&= {\rm sin} \ \theta_1 \ {\rm cos} \ \theta_2 \ ,
\nonumber  \\
&&. 
\nonumber \\
&&.
\nonumber \\
&&.
\nonumber \\
\omega_{D-1} &&= {\rm sin} \ \theta_1 \dots {\rm sin} \ \theta_{D-2} \ {\rm cos} \ \theta_{D-1} \ ,
\nonumber \\
\omega_{D}&& = {\rm sin} \ \theta_1 \dots {\rm sin} \ \theta_{D-2} \ {\rm sin} \ \theta_{D-1} \ ,
\label{parametrization}
\end{eqnarray}
with $0\le \theta_i < \pi$ for $1\le i < D-1$ and $0\le \theta_{D-1} <2\pi$ and metric given by
\begin{equation}
d\Omega^2 = \sum_{i=1}^D d\omega_i^2 = d\theta_1^2 + {\rm sin}^2 \theta_1 d\theta_2^2 + \dots + {\rm sin}^2 \theta_1 \dots {\rm sin} ^2 \theta_{D-2} d\theta_{D-1}^2 \ .
\label{angmetric}
\end{equation} 
The manifold can be obtained by an embedding into a $(D+1)$-dimensional Minkowski space such that 
\begin{eqnarray}
X_0 &&= {1\over H} {\rm cosh} \ Hz \ ,
\nonumber \\
X_i &&= {1\over H} {\rm sinh} \ Hz \ \omega_i \ , \qquad i=1 \dots D \ ,
\label{embedding}
\end{eqnarray}
so that $ds^2 =-dX_0^2 + \sum_{i=1}^D dX_i^2$ and $X_0^2 -\sum_{i=1}^D X_i^2 = 1/H^2$. Note that the manifold is intrinsically Riemannian even if it is embedded in Minkowski space. 

Brownian motion on the universal cover of a general negative-curvature manifold is asymptotically ballistic \cite{ledrappierbal}. Let $B(t)$ be the Brownian motion starting from a point $x$ with coordinate $Hz(x) \gg 1$ and $d$ the distance on the manifold. Then the central limit theorem in negative curvature \cite{ledrappierbal} implies that there exist a positive number $w$ such that
\begin{equation}
{d\left( x, B(t) \right) \over t}=  w + O\left( {1\over \sqrt{t}} \right) {\cal N} (0,1)\ .
\label{asymprad}
\end{equation}
where ${\cal N} (0,1)$ denotes a Gaussian of mean zero and unit standard deviation. This implies that, asymptotically, where this Gaussian distribution can be neglected, Brownian motion on a negative-curvature manifold is invariant under its isometries. In case of constant negative curvature, the group of isometries is ${\it O}_+(D,1)$, or ${\it SO}_+(D,1)$ if we impose preservation of orientations. This follows immediately from the fact that the manifold metric is induced by the embedding in a Minkowski space of one higher dimension. The ``+" subscript, indicates that the Lorentz group is orthochronous, i.e. that the time-reversal symmetry is absent. This is an expression of the fact that a constant negative curvature manifold is one sheet of a generalized two-sheeted hyperboloid: time reversal would connect the two sheets but is not a symmetry of a single sheet. 

On a constant negative-curvature manifold (\ref{radial}), Brownian motion can be decomposed in a radial component $z(t)$ and and angular component $\Omega (t)$ \cite{hsu, hsubrief, arnaudon, hsubook}. The radial part satisfies
\begin{equation}
z(t) = z_0 + W(t)  + {D-1\over 2} \int_0^t {G^\prime (z(s)) \over G(z(s)) } ds \ ,
\label{radialbm1}
\end{equation}
where $W(t)$ is a 1-dimensional Brownian motion \cite{hsubrief, arnaudon} and $t$ is the absolute Parisi-Wu time previously discussed. Using $G(z)$ in (\ref{radial}) we obtain
\begin{equation}
{z(t) \over t} = {r_0 + W(t)\over t} + {D-1\over 2} {H\over t} \int_0^t {\rm coth} \ Hz(s) \ ds \ .
\label{radialbm2}
\end{equation}
Since the standard deviation of the 1D Brownian motion $W(t)$ is $\sigma_{W(t)} \propto \sqrt{t}$ and ${\rm coth} \ Hz > 1 $ for positive $z$, the probabilistic component can be neglected for large $t$ and $z(t) \to \infty$. We thus find the leading-order deterministic asymptotic behaviour \cite{hsubrief}
\begin{equation}
{z(t) \over t} = {D-1 \over 2} H  + O\left( {1\over \sqrt{t}} \right) {\cal N} (0,1)\ , \qquad Hz(t) \gg 1 \ ,
\label{asymprad}
\end{equation}
For test particles reaching the asymptotic region $Hz \gg 1$ the random component becomes negligible and the manifold coordinate $z$ is dynamically soldered with absolute time and emerges as general-relativistic coordinate time. We will for the moment focus on this deterministic leading order and return to the subdominant random component later.

At this point we have to pause to discuss dimensions. The manifold coordinate $z$ has dimension length, as does the quantity $1/H$, which is the radius of curvature. The absolute time $t$ is a diffusion parameter and, as such, has the usual diffusion coefficient ${\cal D}$ with dimension ${\rm length}^2/{\rm time}$ absorbed in its definition, so that $t= {\cal D} \tilde t $ is actually a ${\rm length}^2$, whereas $\tilde t $ is measured in ``seconds". The radius of curvature $1/H$ and the diffusion coefficient ${\cal D}$ can be traded for one length scale and a fundamental speed $c= {\cal D}H$, so that $Ht = c\tilde t$ has dimension length. 

The angular process $\Omega (t)$ converges on the sphere ${\cal S}^{D-1}$. It can be transformed into standard Brownian motion $\Omega (T) $ on ${\cal S}^{D-1}$ by the time change \cite{hsu, arnaudon} 
\begin{equation}
T(t) = \int_0^t  {1\over G\left( z_s \right)^2} \ ,
\label{newtime}
\end{equation}
and its quadratic variation is given by \cite{arnaudon} 
\begin{equation}
[ \Omega^2] (t) = (D-1)  T(t)  \ .
\label{quadvartot}
\end{equation} 
In general, the new time $T(t)$ depends itself on a random variable. However, this is not so asymptotically. Let us consider Brownian trajectories over the time span $[t, \infty)$ that converge to the same angle $\Omega_{\infty}$, where $t \gg 1/H^2$. Because of continuity, their quadratic variation is 
\begin{equation} 
[\Omega_{\infty}^2] (t) = (D-1)  \int_t^\infty ds {1\over G\left( z_s \right)^2} \ .
\label{quadvarinf}
\end{equation} 
Since $H^2t \gg 1$ we can substitute $z(s)$ with its limiting dominant deterministic value (\ref{asymprad}). Then, using (\ref{radial}) we obtain 
\begin{eqnarray}
[ \Omega_{\infty}^2] (z) &&= 4\ {\rm e}^{-2Hz} + O\left( {\rm e}^{-4Hz} \right) 
\nonumber \\
&&= 4\ {\rm e}^{-2H(D-1)w} + O\left( {\rm e}^{-4H(D-1)w} \right) \ ,
\label{limang}
\end{eqnarray}
where we have introduced the new variable $w=Ht/2 = c\tau/2$, such that $z=(D-1)w$ and we will henceforth neglect the subdominant exponential. 

The quadratic variation of Brownian motion satisfies, on large enough scales, the Einstein relation, 
\begin{equation}
[ x^2] (t)  = t \ ,
\label{normaldiff}
\end{equation}
which is equivalent (in 1D)  to the diffusion equation 
\begin{equation}
{\partial \over \partial t} u = {\partial^2 \over \partial x^2} u \ ,
\label{normalheat}
\end{equation}
for the probability distribution $u(t, x)$ of the diffusing particles. On scales shorter than the mean free path $\ell_p = v \tau_p$, however, Brownian motion is ballistic (see \cite{bal} and references therein), with a quadratic variation proportional to the square of time: particles moves along geodesics with constant velocity $v$, 
\begin{equation}
[ x^2](t)  = {1\over \ell_p^2} t^{2} \ .
\label{anomalousdiff}
\end{equation}
In particular, Brownian trajectories originating from a given point always start out as ballistic, up to the characteristic time scale $\tau_p$ \cite{bal}. In the present case, however, we have the reversed situation: since the angular Brownian motion converges to a fixed angle, it is the final section of the converging trajectories that becomes ballistic for $Hz \gg 1$. In particular 
\begin{eqnarray}
[\Omega_{\infty}^2 ](z)  && = \tau^{2} \ ,
\nonumber \\
\tau  = -2\ {\rm e}^{-Hz} &&= -2 \ {\rm e}^{-H(D-1)w} \ ,
\label{anomalousdiff}
\end{eqnarray}
where $\tau $ will be momentarily identified with what is known as conformal time in expansionary cosmology, specifically in de Sitter space-time. Note that small intervals in $\tau$ correspond to large intervals in coordinate time $z$ for $Hz \gg 1$. Essentially, the constant characteristic time of angular Brownian scatterings in the rescaled time $T(t)$ (or $\tau (t)$) gets diluted away in the emerging coordinate time $z$ and only ballistic motion survives. In this asymptotic regime, the trajectories are concentrated on rays $\left[d\left(\Omega(\tau)-\Omega_\infty \right)- \tau \right]$, with $d$ the geodesic distance on the sphere and the probability distribution of particles with coordinates $\theta_i$ and $z\gg 1/H$ is governed by the wave equation 
\begin{equation}
\left( - {\partial^2 \over \partial \tau^2} + \Delta_\Omega \right) u\left(z, \theta_i \right) = 0 \ ,
\label{bal1}
\end{equation}
where $\Delta_\Omega$ is the Laplacian on the sphere. 

We would now like to derive the corresponding wave equation in coordinates $(w, \theta_i)$, instead of $(z, \theta_i)$. To do so we first note that (\ref{bal1}) can be written as
\begin{equation}
\left( - {1\over r_o^2} {\partial^2 \over \partial \tau^2} + {1\over r_o^2} \Delta_\Omega \right) u\left(z, \theta_i\right) = 0 \ ,
\label{bal2}
\end{equation}
where $r_o = (1/2H) \ {\rm exp} Hz$ is the original radius (asymptotically) in (\ref{radial}) and $(1/r_o^2) \Delta_\Omega$, the angular component of the Laplacian, is the generator of angular Brownian motion at fixed radius $r_0$ in embedding space. We are primarily interested, however, in angular Brownian motion at fixed radius $(D-1) r_n$ with $r_n = (1/2H) \ {\rm exp} (Hw)$ in coordinates $w$ and $\theta_i$. This is generated by $(1/(D-1)^2 r_n^2) \Delta_\Omega$. Correspondingly, the wave equation in coordinates $(w, \theta_i)$ becomes
\begin{equation}
\left( - {\partial^2 \over \partial \tau^2} + {1\over (D-1)^2} {r_o^2\over r_n^2} \Delta_\Omega \right) u \left( w, \theta_i \right) = 0 \ ,
\label{angwave}
\end{equation}
where now both $\theta $ and the radii have all to be considered as functions of the new coordinate $w$. 
Expressed through this new coordinate $w$ we obtain
\begin{equation} 
\left( - {\partial^2 \over \partial_w^2} - (D-1) H {\partial \over \partial_w} + 4H^2 {\rm e}^{-2Hw} \Delta_\Omega \right) u\left( w, \theta_i \right) =0.
\label{inw}
\end{equation}
In the asymptotic regime $Hw \gg 1$, this is the wave equation on a Lorentzian manifold of positive curvature $+H^2$ with coordinates $w$ and $\theta_i$, obtained by substituting ${\rm cosh} \leftrightarrow {\rm sinh} $ in (\ref{embedding}). This manifold is exactly de Sitter space-time \cite{strominger}. We have thus derived that Brownian motion on a Riemannian manifold of negative curvature propagates asymptotically along the null geodesics of the corresponding Lorentzian de Sitter space-time, after the absolute diffusion time is dynamically soldered to the radial coordinate of the manifold. The absolute value of the negative curvature becomes the cosmological constant \cite{cosconst}. The original negative-curvature Riemannian manifold is thus asymptotically equivalent to an ``emergent, effective de Sitter space-time". The prescription for switching from the fundamental to the effective description entails reversing the signs of both the time component of the metric and the curvature.

To make contact with our spatially flat universe we shall also consider the flat slicing of the negative-curvature manifold, given by the embedding 
\begin{eqnarray}
X_0 &&= {1\over H} {\rm cosh} Hv +{1\over 2H} r^2 {\rm exp}(Hv) \ ,
\nonumber \\
X_1 &&= {1\over H} {\rm sinh} Hv -{1\over 2H} r^2 {\rm exp}(Hv)
\nonumber \\
X_i &&= {x_i \over H} {\rm e}^{Hv} \ , \qquad i=2 \dots D \ ,
\label{embeddingflat}
\end{eqnarray}
with $r^2 = \sum_i x_i^2$ and the metric taking the form $ds^2 = dv^2 +(1/H^2) {\rm exp}(2Hv) dx^2 $ with $dx^2 = \sum_i dx_i^2$ the flat Euclidean metric on ${\mathbb R}^{D-1}$ (with this definition the $x_i$ are dimensionless, distances are measured in units of $1/H$). 
Since the two times $z$ and $v$ coincide asymptotically we have the same picture: asymptotically, the Lorentzian picture with $v$ as a time coordinate of the flat slicing of de Sitter space-time with ${\rm cosh} \leftrightarrow {\rm sinh} $ in (\ref{embeddingflat}) is an effective description of ballistic diffusion on the fundamental negative-curvature manifold. 

The dynamical emergence of coordinate time and the dilution of Brownian scatterings mean that, asymptotically, the equations describing the original Brownian motion can be formulated entirely as deterministic, ballistic equations in terms of polar manifold coordinates. This, however, implies further that these equations, as viewed by an observer in another inertial frame of reference, should be covariant under the isometry group of the emergent, effective de Sitter space-time while the original absolute time survives as cosmic time. This covariance is satisfied, since the de Sitter isometry group ${\rm SO} (D,1)$ coincides, up to time-reversal symmetry, with the exact ground state isometry group ${\rm SO}_+ (D,1)$. The absence of time-reversal symmetry means that the emergent coordinate time automatically comes endowed with a forward arrow. Note that time-reversal symmetry violation is one of the three Sakharov conditions needed for a matter-antimatter asymmetry in the universe (for a review see \cite{peccei}). The de Sitter group ${\rm SO}(D,1)$ replaces the Poincar\'e group of special relativity when not only a velocity $c$ is fixed but also a length scale $1/H$ \cite{dSrel1} (for a review see \cite{dSrel2}). In particular, the homogenous Lorentz subgroup still describes rotations and boosts, while the translations of the Poincar\'e group become combinations of translations and proper conformal transformations \cite{dSrel3}. The breakdown of ordinary translation symmetry is the only deviation of de Sitter relativity from ordinary special relativity. 

\iffalse

The basic idea of the present model is that, for asymptotic Brownian motion in constant negative curvature, one coordinate emerges dynamically as time, while Brownian scatterings in the orthogonal ``spatial" motion are diluted away so that only the ballistic motion below the characteristic scattering time survives. If this short-time dynamics is generalized to include complex probability amplitudes and quantum walks (for a review see e.g. \cite{quantumwalk}), then the isometry group of the constant negative curvature manifold guarantees that all familiar Lorentz representations are admitted as asymptotic particle excitations, including massive ones. Quantum walks would then describe the microscopic dynamics of matter with respect to the universal time; this is decohered by scatterings on the link fluctuations of the fundamental graph, becoming thus Brownian motion \cite{decoherence}. In the asymptotic region of the manifold where these graph fluctuations and the corresponding Brownian scatterings are diluted away, quantum walks in universal time become the usual quantum behaviour for Lorentz representation with one manifold coordinate as time. 

\fi

Every geodesically complete, D-dimensional manifold $M$ of constant negative curvature has a natural geometric boundary manifold $\partial M$ defined as the locus of all equivalence classes of geodesic rays that remain in bounded distance of each other. If $M$ is simply connected, the boundary $\partial M$ is a sphere $S^{D-1}$ \cite{eberlein} (for a review see e.e. \cite{ratcliffe}). 
Moreover, there is a one-to-one correspondence between interior $SO_{+}(D,1)$ isometries on $M$ and (D-1)-dimensional conformal transformations on the boundary $\partial M$ \cite{ratcliffe}. This correspondence is a Riemannian version of the holographic principle on Lorentzian negative-curvature manifolds like anti-de-Sitter (adS) space \cite{holo1, holo2} (for a review see \cite{holorev}). However, via the emerging time coordinate it defines asymptotically holography on effective de Sitter space-time. 

Let us finally focus on the spectral dimension of effective de Sitter space-time. In general, diffusion processes probe the intrinsic geometry of a manifold. In particular, the heat kernel trace (here $t$ is the same time as originally introduced in the Brownian motion equations (\ref{asymprad}), (\ref{radialbm1}) and following)
\begin{equation}
K(t) = {\rm tr} \ {\rm e}^{\Delta t} \ ,
\label{hkt}
\end{equation}
measures the return probability after time $t$ and defines the spectral dimension of the manifold via the spectral function (for a review see \cite{dunne}) 
\begin{equation}
d_s(t) = -2 {d \ {\rm ln} K(t) \over d\ {\rm ln} t} \ .
\label{specfunc}
\end{equation} 
The spectral function $d_s(t)$ probes the spectral dimension of the manifold in different regimes, on microscopic scales for $t\to 0$, on large scales for $t \to \infty$. It measures the effective number of independent directions available for a random walker. For a Euclidean manifold it coincides with its topological dimension and is independent of $t$. In general, however it can differ from the topological dimension and needs not even be an integer for fractals and graphs. 

The heat trace kernel for manifolds of constant negative curvature $(-H^2)$ has the uniform continuous estimate \cite{davies, grigorian}
\begin{equation}
K(t) \asymp {\left( 1+H^2 t \right)^{(D-3)/2} \over (H^2t)^{D/2}} {\rm e}^{ -{(D-1)^2 \over 4} H^2 t} \ ,
\label{returnCH}
\end{equation}
which gives the spectral curve
\begin{equation}
d_{\rm s} (t)  = D- (D-3) \left( {H^2 t\over 1+H^2 t} \right) +{(D-1)^2\over 2} H^2 t \ .
\label{specCH}
\end{equation}
In general, one would expect that, in the limit $t\to \infty$ of infinite diffusion time, the diffusion process on $M$ probes the ``geometry at infinity". However, in the present case this is not so since the spectral function diverges for $t \to \infty$. This is because, on a negative curvature manifold, the Laplacian has a spectral gap
\begin{equation}
\lambda_0 = - {\rm lim}_{t\to \infty} {{\rm ln}\ K(t) \over t} = {(D-1)^2H^2\over 4}  \ ,
\label{spegap}
\end{equation} 
representing the bottom of the spectrum of the positive operator $-\Delta$. As a consequence of this spectral gap, the return probabilities $K(t)$ are dominated by an exponential behaviour at large $t$, which is equivalent to ballistic diffusion. In other words, because of ballistic diffusion, the process runs away to infinity too quickly to sample the geometric properties of the manifold \cite{anker}. 

If one wants to probe the geometry on large scales one must define a modified, slower diffusion process, dominated by the Laplacian eigenvalues just above $\lambda_0$ after the spectral gap, i.e. the runaway deterministic time dimension has been subtracted. 
This can be done and this slower diffusion process is called the infinite Brownian loop \cite{anker}. The infinite Brownian loop is the infinite-$\Theta$ limit of the Brownian bridge $BB(t,\Theta)$, which is the Brownian motion $B(t)$ constrained to come back to the origin $x=0$ for $t = \Theta$, 
\begin{equation}
B(t) =  {t\over \Theta} B(\Theta) + BB(t,\Theta) \ .
\label{bridge}
\end{equation} 
When Brownian motion is not asymptotically ballistic, the infinite Brownian loop is the Brownian process itself. Otherwise, it represents the subdominant random component after the dominant deterministic component has been subtracted. 

Let us denote by $\varphi_0$ the most symmetric eigenstate of the Laplacian corresponding to the lowest eigenvalue $\lambda_0$, $\left( \Delta + \lambda_0 \right) \varphi_0 = 0$. The infinite Brownian loop is then the relativized $\varphi_0$-process \cite{sullivan}, with generator
\begin{equation}
\tilde \Delta (f) = {1\over \varphi_0} \Delta \left( f\varphi_0 \right) + \lambda_0 f = \tilde \Delta f + 2 \nabla {\rm ln} \ \varphi_0 \cdot \nabla f \ .
\label{relproc}
\end{equation}
As anticipated, in this process measured relatively to the ground state $\varphi_0$, the spectral gap falls out. This corresponds to dropping the exponential in (\ref{returnCH}), which gives the modified spectral function 
\begin{equation} 
\tilde d_{\rm s} (t)  = D- (D-3) \left( {H^2 t\over 1+H^2 t} \right) \ ,
\label{modspe}
\end{equation}
whose infinite-time limit is 
\begin{equation}
D_{\rm inf} = {\rm lim}_{t \to \infty} \ \tilde   d_s(t) = 3 \ ,
\label{dimthree}
\end{equation}
independently of $D$. The quantity $D_{\rm inf}$ is called the ``pseudo-dimension", or ``dimension at infinity" of a constant negative curvature manifold \cite{anker}. It measures the asymptotic spectral dimension of the manifold on large scales, as probed by the slow random process in the radial direction which survives after ``subtracting deterministic time". This slow random process always sees three Euclidean dimensions and it is confined to the Weyl chamber \cite{anker}, which means, in this case, that it is confined to the forward direction, another expression of the emergent arrow of time. 

We have derived the spatial spectral dimension three on large scales in the case of constant negative curvature. This result, however is a general consequence of the local limit theorem \cite{ledrappier} and is valid for any manifold of strictly negative curvature. This theorem states that, on the universal cover of any closed and connected manifold of strictly negative curvature, there exists a positive constant $C$ such that 
\begin{equation}
{\rm lim}_{t\to \infty} \ t^{3/2} {\rm e}^{\lambda_0 t} K(t)  = C \ ,
\label{llt}
\end{equation} 
where $K(t)$ is the heat kernel trace and $\lambda_0$ is the spectral gap. In other words, when the spectral gap (time) is ``subtracted" the spectral dimension at infinity is always three, independently of the topological dimension $D$. 

On effective de Sitter space-time the spectral dimension changes thus from $D$ Riemannian dimensions and a universal time on microscopic graph scales to (3+1) Lorentzian dimensions on large scales. Perhaps the most interesting case is $D=2$. In this case the cosmological ground state is a fluctuating negative-curvature surface (note that this is a well defined surface with Hausdorff dimension 2 \cite{comb4}) and the flow from 2 Euclidean spectral dimensions to (3+1) Lorentzian dimension is reminiscent of causal dynamical triangulations \cite{cdt}, with the difference that here coordinate time and manifold causality are emergent. 

Let us finally focus on the properties of the sub-dominant random process for constant negative curvature. In this case, the uniform continuous estimate of the kernel of the infinite Brownian loop is known \cite{davies}. For large $t$ it is given by
\begin{equation}
K(t, \rho) \asymp t^{-{3\over 2}} {\rm e^{-{\rho^2 \over 4t}}} \ ,
\label{iblkernel}
\end{equation}
where $\rho$ is the hyperbolic distance. This is the isotropic heat kernel on a 3D Euclidean manifold with norm given by $\rho$ \cite{anker}. To obtain the effective description of the infinite Brownian loop we have to switch the signs of the time component of the metric and of the curvature. In this Euclidean case, this amounts to the the familiar Wick rotation. Because of the leading-order soldering in eq. (\ref{asymprad}), also the absolute time $t$ must be rotated to obtain the effective kernel in the asymptotic region, 
\begin{equation}
K_{\rm eff} (\tilde t, \rho) \asymp \left( {m\over i\hbar \tilde t} \right)^{3\over 2} \  {\rm e^{-{m \rho^2 \over 2i\hbar  \tilde t}}} \ ,
\label{qm}
\end{equation}
where we have used the previously introduced time $\tilde t$ measured in seconds. This is the 3D Euclidean Schrödinger propagator for a particle of mass
\begin{equation}
m={\hbar H\over 2c} \ ,
\label{mass}
\end{equation}
whose value reflects the Einstein equation relating energy density to curvature.

\section{Conclusion}
The picture emerging from the present model is that the asymptotic appearance of de Sitter space-time with a cosmological constant is only an effective description of a different fundamental ground state. Going backwards in time, ever more frequent stochastic scatterings with the fundamental graph fluctuations turn the coordinate time of de Sitter space-time into a random variable on a Riemannian negative curvature manifold. There are no singularities, no horizon and the holographic principle appears naturally due to the fundamental negative curvature. The spectral dimension of the resulting asymptotic manifold is (3+1) independent of its topological dimension and quantum behaviour appears as the effective description of the sub-dominant surviving random process.

\section{Acknowledgments}
We thank M. Arnaudon, J.-H. Eschenburg P. Bougerol and F. Ledrappier for extremely helpful discussions.


\begin{thebibliography}{10}
	\expandafter\ifx\csname url\endcsname\relax
	\def\url#1{\texttt{#1}}\fi
	\expandafter\ifx\csname urlprefix\endcsname\relax\def\urlprefix{URL }\fi
	\providecommand{\bibinfo}[2]{#2}
	\providecommand{\eprint}[2][]{\url{#2}}
	
	
\bibitem{strominger} M. Spradlin, A. Strominger and A. Volovich, Les Houches lectures on de Sitter space, arXiv:hep-th/0110007. 
	
%\bibitem{weinberg} S. Weinberg, Cosmology, Oxford University Press, Oxford (2008). 

%\bibitem{rugh} S. E. Rugh and H. Zinkernagel, On the physical basis of cosmic time, {\it Stud. His. Philos. Mod. Phys. } {\bf 40} 1-19 (2009).
	
%\bibitem{strings} J. Polchinski, String Theory, Cambridge University Press, Cambridge (1998). 
	
%\bibitem{as} A. Eichhorn, Asymptotically safe gravity, proceedings of the 57th Course of the Erice International School of Subnuclear Physics, "In search for the unexpected", June 2019, arXiv: 2003.00044, (2019). 

%\bibitem{cdt1} J Ambjorn, J. G\"orlich, J. Jurkiewicz and R. Loll, Nonperturbative Quantum Gravity, {\it Phys. Rep. } {\bf 519} 127-210 (2012) 

%\bibitem{cdt2} R. Loll, Quantum Gravity from Causal Dynamical Triangulations: A Review, {\it Classical and Quantum Gravity}  {\bf 37} 013002 (2019). 
	
%\bibitem{causalsets} S. Surya, The causal set approach to quantum gravity, {\it Living Reviews in Relativity} {\bf 22:5} (2019). 

\bibitem{susskind1} N. Goheer, M. Kleban and L. Susskind, The trouble with de Sitter space, {\it JHEP} {\bf 07} 056 (2003). 

\bibitem{dine} M. Dine, J. A. P. Law-Smith and S. Sun, Obstacles to construct de Sitter space in string theory, {\it JHEP} {\bf 50} (2021). 

\bibitem{kuchar} K. V. Kuchar, Time and interpretations of quantum gravity, {\it Int. J. Mod. Phys.} {\bf D20} 3 (2011). 

%\bibitem{stuckelberg}E. C. G. St\"uckelberg, Remarks on the creation of pairs of particles in the theory of relativity, {\it Helv. Phys. Acta} {\bf 14} 588-594 (1941).

%\bibitem{horwitz} L. P. Horwitz and C. Piron, Relativistic dynamics, {\it Helv. Phys. Acta} {\bf 46} 316-326 (1973). 

%\bibitem{abbot} L. F. Abbot and S. Deser, Stability of gravity with a cosmological constant, {\it Nucl. Phys.} {\bf B195} 76-96 (182). 

%\bibitem{polyakov} A. Polyakov, De Sitter space and eternity, {\it Nucl. Phys.} {\bf 797} 199-217 (2008). 

%\bibitem{susskind2} L. Susskind, De Sitter holography: fluctuations, anomalous symmetry and wormholes, {\it Universe} {\bf 7} 464 (2021). 

%\bibitem{hh}J. Hartle and J. Hawking, Wave function of the universe, {\it Phys. Rev.} {\bf 28} 2960 (1983). 

%\bibitem{hhrev} J. Hartle, The quantum universe, World Scientific, Singapore (2021). 

\bibitem{comb1} C. A. Trugenberger, Combinatorial quantum gravity: geometry from random bits,  {\it JHEP} {\bf 09} 045 (2017).

\bibitem{comb2} C. Kelly, C. A. Trugenberger, and F. Biancalana, Self-Assembly of Geometric Space from Random Graphs, {\it Classical and Quantum Gravity} {\bf 36} 125012 (2019). 

\bibitem{comb3} C. Kelly, C. A. Trugenberger and F. Biancalana, Emergence of the circle in a statistical model of random cubic graphs, {\it Classical and Quantum Gravity} {\bf 38} 075008 (2021). 

\bibitem{comb4} C. A. Trugenberger, Emergent time, cosmological constant and boundary dimension at infinity in combinatorial quantum gravity, {\it JHEP} {\bf 19} (2022). 

\bibitem{olli0} Y. Ollivier, Ricci curvature of metric spaces, {\it C. R. Math. Acad. Sci. Paris} {\bf 345} 643-646 (2007). 

\bibitem{olli1}Y. Ollivier, Ricci curvature of Markov chains in metric spaces, {\it J. Funct. Anal.} {\bf 256} (2009) 810; 

\bibitem{olli2} Y. Ollivier, A survey of Ricci curvature fo metric spaces and Markov chains, {\it Adv. Stud. Pure Math.} {\bf 57} (2010) 343-381 (2010). 

\bibitem{olli3}Y. Linn, L. Lu and S. T. Yau, Ricci curvature of graphs, {\it Tohoku Math. J.} {\bf 63} (2011) 605-627. 

\bibitem{olli4} J. Jost and S. Liu, Ollivier's Ricci curvature, local clustering and curvature dimension inequalities on graphs, {\it Discrete Comput. Geom.} {\bf 51} (2014) 300-322. 

\bibitem{wheeler} J. A. Wheeler, Information, physics, quantum: the search for links, Proceedings of the III international symposium on the foundations of quantum mechanics, 354-358, Tokyo (1989) . 

\bibitem{deutsch} D. Deutsch, It from Qubit, in Science and ultimate reality, J. Barrow, P. Davies and C. Harper (eds.) Cambridge University Press, Cambridge (2003).

\bibitem{shiga} K. Shiga, Hadamard manifolds, in ``Geometry of geodesics and related topics", {\it Advanced Studies in Pure Mathematics} {\bf 3} 239-281 (1984). 

%\bibitem{stochastic} P. H. Damgaard and H. H\"uffel, Stochastic quantization. {\it Phys. Rep.} {\bf 152} 227-398 (1987). 

%\bibitem{parisiwu}G. Parisi nd Y.-S. Wu, {\it Sci. Sinica} {\bf 24} 483 (1981). 

\bibitem{pessa} A. B. Pessa and H. R. Ribeiro, Characterizing stochastic time series with ordinal networks, {\it Phys. Rev.} {\bf E100} 042304 (2019). 

\bibitem{eastman} P. Eastman, Introduction to statistical mechanics. Stanford University (2014). 

\bibitem{kleinertcur} H. Kleinert and S. V. Shabanov, Theory of Brownian motion of a massive particle in spaces with curvature and torsion, {\it J. Phys. A: Math. Gen.} {\bf 31} 7005-7009 (1998). 

\bibitem{ledrappierbal} F. Ledrappier, Central limit theorem in negative curvature, {\it The Annals of Porbability} {\bf 3} 1219-1233 (1995). 
	
\bibitem{hsu} P. Hsu and W. S. Kendall, Limiting angle of Brownian motion in certain two-dimensional Cartan-Hadamard manifolds {\it Annales de la facult\'e des Sciences de Toulouse} {\bf 1} 169-186 (1982). 

\bibitem{hsubrief} E. P. Hsu, A brief introduction to Brownian motion on a Riemann manifold, Summer School in Kyushu (2008). 

\bibitem{arnaudon}M.Arnaudon and A. Thalmeier, Brownian motion and negative curvature, {\it Progress in Probability} {\bf 64} 145-163 (2011). 

\bibitem{hsubook} E. P. Hsu, Stochastic analysis on manifolds, {\it Graduate studies in mathematics} {\bf 38}, Providence (RI) (2002). 	

\bibitem{dSrel1} R. Aldrovandi, J. P. Beltran Almeida and J. G. Pereira, de Sitter special relativity, {\it Class. Quant. Grav.} {\bf 24} 1385-1407 (2007).

\bibitem{dSrel2} R. Aldrovandi and J. G. Pereira, de Sitter relativity: a new road to quantum gravity?, {\it Foundations of Physics} {\bf 39} 1-19 (2009). 

\bibitem{dSrel3} R. Aldrovandi, J. P. Beltran Almeida, C. S, Mayor and J. G. Pereira, Lorentz transformations in de Sitter relativity, arXiv:0709.3947 (2007). 
	
\bibitem{anker} J.-P. Anker, P. Bougerol and T. Jeulin, The infinite Brownian loop on a symmetric space, {\it Rev. Mat. Iberoamericana} {\bf 18} 41-97 (2002). 

\bibitem{ledrappier} F. Ledrappier and S. Lim, Local limit theorem in negative curvature, arXiv:1503.04156 (2020). 

\bibitem{hh}J. Hartle and J. Hawking, Wave function of the universe, {\it Phys. Rev.} {\bf 28} 2960 (1983). 

\bibitem{hhrev} J. Hartle, The quantum universe, World Scientific, Singapore (2021). 

%\bibitem{ledrappier}F. Ledrappier, Applications of dynamics to compact manifolds of negative curvature, in {\it Proceedings of the international congress of mathematiciansy, Z\"urich}, S. D. Chatterji ed., 1195-1202, Springer-Verlag (1994). 

%\bibitem{isham} C. J. Isham, Canonical Quantum Gravity and the Problem of Time. In: Integrable Systems, Quantum Groups, and Quantum Field Theories, L. A. Ibort, M. A. Rodriguez (eds.), NATO ASI Series (Series C: Mathematical and Physical Sciences), vol 409. Springer-Verlag, Berlin (1993). 
	
%\bibitem{page}D. N. Page and W. K. Wooters, Evolution without evolution: dynamics described by stationary observables, {\it Phys. Rev. }{y\bf D27} 2885, (1983). 
	
%\bibitem{raamsdonk1}M. V. Raamsdonk, Buidling up spacetime with quantum entanglement, {\it Gen. Rel. Grav.} {\bf 42} 2323-2329 (2010). 	

%\bibitem{carroll1} C. Cao, S. M. Carroll and S. Michalakis, Space from Hilbert space: recovering geometry from bulk entanglement, {\it Phys. Rev. } {\bf D95} 024031 (2016). 

%\bibitem{raamsdonk2}M. V. Raamsdonk, Building up spacetime with quantum entanglement II: it from BC-bit, arXiv:1809.01197.
	
%\bibitem{carroll2} C. Cao and S. M. Carroll, Bulk entanglement gravity without a boundary: towards finding Einstein's equation in Hilbert space, {\it Phys. Rev. } {\bf D97} 086003 (2108). 

I %\bibitem{deutsch}D. Deutsch, It from Qubit, in {\it Science and ultimate reality}, J. Barrow, P. Davies and C. Harper (eds.) Cambridge University Press, Cambridge (2003). 
	
%\bibitem{graphrev}R. Albert and L. Barabasi, Statistical mechanics of complex networks, {\it Rev. Mod. Phys.} {\bf 74} (2002) 47. 
	

%\bibitem{conv1} P. van der Hoorn , W. J. Cunningham, G. Lippner, C. A. Trugenberger and D. Krioukov, Ollivier-Ricci curvature convergence in random geometric graphs, {\it Phys. Rev. Res.} {\bf 3} 013211 (2021). 

%\bibitem{conv2} C. Kelly, C. A. Trugenberger and F. Biancalana, Convergence of combinatorial gravity, arXiv:2102.02356. 
	
%\bibitem{zanardi} S. Garneroni, P. Giorda and P. Zanardi, 	Bipartite quantum states and random complex networks, {\it New Journal of Physics} {\bf 14} 013011 (2012).

%\bibitem{ent} M. C. Diamantini and C. A. Trugenberger, Topological network entanglement as order parameter for the emergence of geometry, {\it New Journal of Physics} {\bf 19} 106741 (2017). 

\bibitem{bal} P. N. Pusey, Brownian motion goes ballistic. {\it Science} {\bf 332} 802 (2011). 


%\bibitem{sandev} T. Sandev, R. Metzler and A. Chechkin, Generalized diffusion and wave equations: recent advances, in Proceedings of the 9th international workshop on {\it Analytical methods of analysis and differential equations}, Cambridge Scientific Publishers (2019). 

%\bibitem{procaccia}V. Ilyin, I. Procaccia and A. Zagorodny, Stochastic processes crossing from ballistic to fractional diffusion with memory: exact results, {\it Condensed Matter Physics} {\bf 13} 23001:1-8 (2010). 

\bibitem{cosconst} T. Padmanabhan, Cosmological constant-the weight of the vacuum, {\it Phys. Rept.} {\bf 380} 235-320 (2003). 

%\bibitem{wavedes}E. L. Lokas, Positive frequency solutions of the Klein-Gordon equation in the N-dimensional de SItter spacetime, {\it Acta Physica Polonica B} {\bf 26} 19 (1995). 

\bibitem{peccei} R. D. Peccei, Matter-antimatter asymmetry in the universe and an arrow for time. arXiv:hep-ph/0608226. 

%\bibitem{quantumwalk} S. E. Venegas-Andraca. Quantum walks: a comprehensive review. {\it Quantum Information Processing} {\bf 11} 1015-1106 (2012). 

%\bibitem{decoherence} A. Romanelli, R. Siri, G. Abal, A. Auyuanet and R. Donangelo, Decoherence in the quantum walk on the line. {\it Physica A: Statistical Mechanics and its Applications} {\bf 357C} (2005). 

\bibitem{levy1} J. Masoliver, Telegraphic transport processes and their fractional generalization: a review and some extensions. {\it Entropy} {\bf 23} 364 (2021). 

\bibitem{levy2} V. Zaburdaev, S. Denisov and J. Klafter, Levy walks. {\it Rev. Mod. Phys.} {\bf 87} 483 (2015). 

\bibitem{eberlein} P. Eberlein and B. O'Neill, Visibility manifolds, {\it Pacific J. Math} {\bf 46} 45-109 (1973). 

\bibitem{ratcliffe} J. G. Ratcliffe, Foundations of hyperbolic manifolds, Springer-Verlag, Berlin (2006). 

\bibitem{holo1} G. 't Hooft, Dimensional Reduction in Quantum Gravity, arXiv:gr-qc/9310026 (1993). 

\bibitem{holo2} L. Susskind, The world as a hologram, {\it J. Math. Phys.} {\bf 36} 6377-6396 (1995). 

\bibitem{holorev} R. Bousso, The holographic principle, {\it Rev. Mod. Phys.} {\bf 74} 835-874 (2002). 
	
\bibitem{dunne} G. Dunne, Heat kernels and zeta functions on fractals, {\it J. Phys.} {\bf A45} 374016 (2012).

\bibitem{davies} E. B. Davies and N. Mandouvalis, Heat Kernel bounds on hyperbolic space and Kleinian groups, {\it Proc. London Math. Soc. (3) } {\bf 52} 182-208 (1988).

\bibitem{grigorian} A. Grigor\' yan, Estimates of heat kernels on Riemannian manifolds, in ``Spectral Theory and Geometry", Cambridge University Press, Cambridge (2010). 

\bibitem{sullivan} D. Sullivan, Related aspects of positivity in Riemannian geometry, {\it J. Differential Geometry} {\bf 25} 327-351 (1987).

\bibitem{cdt}J. Ambjorn, J.Jurkiewicz and R. Loll, The spectral dimension of the universe is scale dependent. {\it Phys. Rev. Lett.} {\bf 95} 171301 (2005). 



%\bibitem{marias} M. Marias, Eigenfunctions of the Laplacian on rotationally symmetric manifolds, {\it Transactions of the American Mathematical Society} {\bf 350} 4367-4375 (1998). 
	


%\bibitem{causalsets} S. Surya, The causal set approach to quantum gravity, {\it Living Reviews in Relativity} {\bf 22:5} (2019). 


%\bibitem{loopqg} A. Ashtekar and E. Bianchi, A short review of loop quantum gravity, {\it Rep. Prog. Phys} {\bf 84} 042001 (2021). 

%\bibitem{cdt1} 
%J Ambjorn, J. G\"orlich, J. Jurkiewicz and R. Loll, Nonperturbative Quantum Gravity, {\it Phys. Rep. } {\bf 519} 127-210 (2012) 

%\bibitem{cdt2} R. Loll, Quantum Gravity from Causal Dynamical Triangulations: A Review, {\it Classical and Quantum Gravity}  {\bf 37} 013002 (2019). 

%\bibitem{tensor1} J. Ambjorn, B. Durhuus and T. Jonsson, Three-dimensional simplicial quantum grvaity and generalized matrix models, {\it Mod. Phys. Lett.} {\bf A06} 1133-1146 (1991).

%\bibitem{tensor2} N. Sasakura, Tensor model for gravity and orientability of manifold, {\it Mod. Phys. Lett.} {\bf A06} 2613-2624 (1991). 

%\bibitem{sasakura} N. Sasakura, Phase profile of the wave function of canonical tensor model and emergence of large spacetimes, arXiv:2104.11845. 


%\bibitem{polyakov} A. Polyakov, Two-dimensional quantum gravity, superconductivity at high $T_c$ in {\it Fields, strings and critical phenomena}, Les Houches 1988, E. Br\`ezin and J. Zinn-Justin eds., North-Holland, Amsterdam (1990). 

%\bibitem{tt} A. Connes and C. Rovelli, Von Neumann algebra automorphisms and time-thermodynamics relation in general covariant quantum theories, {\it Class. Quant. Grav.} {\bf 11} 2899-2918 (1994). 

%\bibitem{tee} P. Tee and C. A. Trugenberger, Enhanced Forman curvature and its relation to Ollivier curvature, {\it EPL} {\bf 133} 60006 (2021). 

%\bibitem{forman} R. Forman, Combinatorial Morse theory, {\it Int. J. Math} {\bf 13} 333-368  (2002).

%\bibitem{klit1} N. Klitgaard and R. Loll, Introducing Quantum Ricci Curvature, {\it Phys. Rev.}  {\bf D97} 046008 (2018). 

%\bibitem{klit2} N. Klitgaard and R. Loll, Implementing Quantum Ricci Curvature, {\it Phys. Rev.} {\bf D97} 106017 (2018). 

%\bibitem{klit3} N. Klitgaard and R. Loll, How round is the quantum de Sitter universe? {\it Eur. Phys. J.} {\bf 80} 990 (2020). 


%\bibitem{gorsky}A. Gorsky and O. Valba, Interacting thermofield doubles and critical behaviour in random regular graphs, {\it Phys. Rev.} {\bf D103} 106013 (2021). 

%\bibitem{dall}J. Dall and M. Christensen, Random geometric graphs, {\it Phys. Rev. } {\bf E66} (2002) 016121.

%\bibitem{krioukov} D. Krioukov, Clustering implies geometry in networtks, {\it Phys. Rev. Lett.} {\bf 116} (2016) 208302. 

%\bibitem{gosztolai}A. Gosztolai and A. Arnaudon, Unfolding the multiscale structure of networks with dynamical Ollivier-Ricci curvature, {\it Nature Communications} {\bf 12} 4561 (2021). 


%\bibitem{lollqf} J. Brunekreef and R. Loll, Quantum flatness in two-dimensional CDT quantum gravity, arXiv:2110.11100. 



%\bibitem{prat} J.-J. Prat. Etude asymptotique et convergence angulaire du mouvement brownien sur une vari\'et\'e \`a courbure n\'egative. {\it C. R. Acad. Sci. Paris S\'er. A-B} {\bf 280(22):Aiii} A1539-A1542 (1975). 

%\bibitem{kendall} W. S. Kendall, Brownian motion on 2-dimensional manifolds of negative curvature. {\it Trans. Amer. Math. Soc.} {\bf 275} 751-760 (1983). 


%\bibitem{chandrashekar} C. M. Chandrashekar, Discrete-time quantum walk-Dynamics and Applications, arXiv:1001.5326. 


%\bibitem{krioukovdesitter} D. Krioukov, M. Kitsak, R. S. Sinkovits, D. Rideout, D. Meyer and M. Boguna, Network Cosmology, {\it Scientific Reports} {\bf 2} 793 (2012). 

\iffalse


\bibitem{genus} C. Thomassen, Embeddings of graphs, {\it Discrete Mathematics} {\bf 124} 217-228 (1994). 

\bibitem{expander}H. Namazi, P. Pankka and J. Souto, Distributional limits of Riemannian manifolds and graphs with sublinear genus growth, {\it Geometric and Functional Analysis} {\bf 24} 322-359 (2014). 

\bibitem{poincare} J. W. Anderson, Hyperbolic geometry, Springer-Verlag, Berlin (2005). 

\bibitem{datta} B. Datta and S. Gupta, Semi-regular tilings of the hyperbolic plane, {\it Discrete \& Computational Geometry} {\bf 65} 531-553 (2021). 
	



\bibitem{kubo}A. J. Maida and P. R. Kramer, Simplified models for turbulent diffusion: theory, numerical modelling and physical phenomena, {\it Phys. Rep.} {\bf 34} 237-574 (1999). 


\bibitem{mcmullen} C. McMullen, Riemann surfaces and the geometrization of 3-manifolds, {\it Bulletin of the American mathematical society} {\bf 27} 207-216 (1992). 

\bibitem{jacobi} J.-P. Anker and V. Pierfelice, Wave and Klein-Gordon equations on hyperbolic spaces, {\it Anal. PDE} {\bf 7} 953-995 (2014). 



\bibitem{cllt} F. Ledrappier and S. Lim, Local limit theorem in negative curvature, arXiv:1503.04156.

\bibitem{three} J.-P. Anker, S. Meda, V. Pierfelice, M. Vallarino and H.-W. Zhang, Schr\"odinger equation on noncompact symmetric spaces, arXiv:2104.00265. 

\fi

%\bibitem{thooft}G. 't Hooft, Dimensional reduction in quantum gravity, arXiv:gr-qc/9310026.

%\bibitem{bousso} R. Bousso, The holograpic principle, {\it Rev. Mod. Phys.} {\bf 74} 825-874 (2002). 

	
	
\end{thebibliography}
\end{document}